 \documentclass[final,5p,times,twocolumn]{elsarticle}

\usepackage{amsmath,amssymb,amsthm,mathtools}
\usepackage{bm}
\usepackage{braket}
\usepackage{graphicx}
\usepackage{physics}
\usepackage{hyperref}
\usepackage{microtype}
\usepackage{doi}
\usepackage[utf8]{inputenc}
\usepackage[german]{babel}

\usepackage{xcolor}

\newtheorem{theorem}{Theorem}[section]

\newtheorem{lemma}[theorem]{Lemma}

\usepackage{bm}

\journal{Physics Letters A}

\begin{document}

\begin{frontmatter}

\title{Nontrivial bounds on extractable energy in quantum energy teleportation for gapped manybody systems with a unique ground state}

\author{Taisanul Haque}
\ead{taisanul.haque@stud.uni-goettingen.de}  
\affiliation{organization={Institute for Theoretical Physics, University of G\"ottingen},
            city={G\"ottingen},
            postcode={37077}, 
            country={Germany}}

\begin{abstract}
We establish {an exponentially decaying} upper bound on the average energy that can be extracted in quantum energy teleportation (QET) protocols executed on finite-range {gapped} lattice systems possessing a unique ground state. Under mild regularity assumptions on the Hamiltonian and uniform operator-norm bounds on the local measurement operators, there exist positive constants $C$ and $\mu$ (determined by the spectral gap, interaction range and local operator norms) such that for any local measurement performed in a region $A$ and any outcome-dependent local unitaries implemented in a disjoint region $B$ separated by distance $d=\operatorname{dist}(A,B)$ one has $|E_A-E_B|\le C\,e^{-\mu d}$.
The bound is nonperturbative, explicit up to model-dependent constants, and follows from the variational characterization of the ground state combined with exponential clustering implied by the spectral gap. {We emphasize that the constants deteriorate as the gap closes (equivalently, as the correlation length diverges), so the estimate is intended for the gapped regime.}
\end{abstract}

\end{frontmatter}

\section{Introduction}
The interplay between quantum information and thermodynamics has unveiled non-trivial protocols that challenge classical intuitions regarding energy transport. Foremost among these is Quantum Energy Teleportation (QET), a theoretical framework originally proposed by Hotta \cite{Hotta2008a,Hotta2008b,Hotta2010}. QET protocols enable the extraction of locally inaccessible energy at a remote site through a sequence of local operations and classical communication (LOCC) performed on an entangled many-body ground state. While a system in its ground state $|G\rangle$ is locally passive---meaning no energy can be extracted by unitary operations acting on a single subsystem---the ground state contains spatial entanglement. {This setting connects to strong local passivity and its refinements, which constrain local energy extraction in finite and many-body systems \cite{FreyFunoHotta2014,Alhambra2019}.} In a QET protocol, a sender (Alice) performs a local measurement on her subsystem, inevitably injecting energy $E_A$ and breaking the ground-state entanglement. By transmitting the measurement outcome to a receiver (Bob) via a classical channel, Bob can utilize the shared correlations to perform a conditional local operation that extracts energy $E_{\mathrm{ext}}>0$. {Operationally, Bob's decoding must be applied on timescales shorter than the characteristic propagation time of energy or excitations between $A$ and $B$, otherwise the protocol reduces to ordinary energy transfer.} This process effectively teleports a fraction of the injected energy across the system without physical energy carriers traversing the intermediate space \cite{Hotta2011}.

Since the first constructive proposals, QET has been investigated in a variety of concrete physical models. These include detailed analyses of spin chains \cite{Hotta2009a}, in phase space quantum mechanics \cite{Sanchez2024}, harmonic lattices \cite{Nambu2010}, equilibrium and nonequilibrium settings \cite{Yan2025}, and trapped ion architectures \cite{Hotta2009b}, as well as proposals for condensed-matter realizations such as quantum Hall systems \cite{Yusa_2011}. Recent experimental breakthroughs have also begun to verify these predictions in controlled quantum platforms \cite{Rodriguez2023, Ikeda2023}. {For recent overviews and broader applications of QET (including algorithmic cooling, negative-energy engineering and multi-agent formulations), see \cite{Ragula2025,RodriguezBriones2017,Funai2017,IkedaSSRN2025}.}

Most literature to date focuses on the optimization of particular measurement/operation pairs to maximize energy transfer in specific Hamiltonians. Here we address the complementary question: \emph{How large can the extractable energy be, in principle, for arbitrary local POVMs and arbitrary outcome-dependent local unitaries?} We answer by proving an absolute upper bound that decays exponentially with the separation between measurement and extraction regions for finite-range gapped lattice Hamiltonians with a unique ground state. {Moreover, our result should be treated strictly in the thermodynamic limit.} 

Our proof uses only elementary operator identities together with two rigorous structural inputs for gapped systems: finite-range locality of the Hamiltonian and exponential clustering of ground-state correlations. The latter follows from Lieb--Robinson bounds combined with the existence of a spectral gap \cite{HastingsKoma2006, NachtergaeleSims2006}. The resulting bound is nonperturbative and explicit up to constants which are in principle computable from the spectral gap, local interaction strength, and the geometric data of the regions involved.

{There are also upper bounds of a different nature that hold without any assumption of a gap or locality. Hotta derived an entanglement-based upper bound for general spin-chain models \cite{Hotta2013}. More recently, Ikeda derived information-theoretic bounds on teleported observables (including energy) that apply to arbitrary Hamiltonians and remain finite in general \cite{Ikeda2025PTEP}. These bounds can be numerically tighter at moderate distances because our prefactor contains boundary sums and measurement-operator norm sums. In contrast, our result gives an explicit exponential dependence on geometric separation in the gapped regime, which is not directly visible in purely information-theoretic inequalities.}

\section{General Framework and Protocol}\label{sec:framework}

We consider a quantum many-body system defined on a discrete lattice $\Lambda$ equipped with a metric $d(\cdot, \cdot)$. The dynamics are governed by a finite-range Hamiltonian $H = \sum_{Z \subset \Lambda} h_Z$, where each local term $h_Z$ acts non-trivially only on the subset of sites $Z$ and has bounded norm $\|h_Z\| \leq J$. We assume the interaction has a finite range {$R$}, such that $h_Z = 0$ if the diameter of $Z$ exceeds {$R$}. Let $|G\rangle$ denote the unique ground state of $H$ with ground-state energy $E_0$. Without loss of generality, we set $E_0 = 0$, so that $H|G\rangle = 0$ and $H \geq 0$.

The spatial separation is defined by two disjoint subregions, $A$ (Alice) and $B$ (Bob), separated by a distance $d = \operatorname{dist}(A,B):=\min_{i \in A, j \in B} d(i, j)$. The standard QET protocol proceeds in three steps:

\begin{enumerate}
	\item \textbf{Measurement:} Alice performs a local general quantum measurement (POVM) on region $A$. This is described by a set of Kraus operators $\{M_k^{A}\}$ satisfying the completeness relation $\sum_k (M_k^{A})^\dagger M_k^{A} = \mathbb{I}_A$. The probability of obtaining outcome $k$ is $p_k = \langle G | (M_k^{A})^\dagger M_k^{A} | G \rangle$. The post-measurement state, conditional on outcome $k$, is $|\psi_k\rangle = \frac{1}{\sqrt{p_k}} M_k^{A} |G\rangle$.

	\item \textbf{Communication:} Alice transmits the classical measurement outcome $k$ to Bob. {Note that this step does not transport energy but allows Bob to select a conditional operation. Bob can only apply his conditional unitary after receiving the classical message. To exclude ordinary energy transport, we assume the classical communication and Bob's local operation are completed before any energy-carrying disturbance created near $A$ can reach $B$.}

	\item \textbf{Extraction:} Based on the outcome $k$, Bob performs a local unitary operation $U_k^{B}$ acting strictly on region $B$. The final state of the system is given by the ensemble $\{ p_k, U_k^{B} |\psi_k\rangle \}$.
\end{enumerate}

The average energy of the system after the execution of the protocol is
\begin{equation}
	E_{\text{fin}} = \sum_k p_k \langle \psi_k | (U_k^{B})^\dagger H U_k^{B} | \psi_k \rangle = \sum_k \langle G | (M_k^{A})^\dagger (U_k^{B})^\dagger H U_k^{B} M_k^{A} | G \rangle.
\end{equation}

The extracted (teleported) energy is defined as the net reduction of the system energy due to Bob's conditional operation, relative to the post-measurement state,
\begin{equation}\label{eq:DefExtEnergy}E_{\mathrm{ext}}:=\Delta E:=E_A-E_B\ge 0.
\end{equation}
{Explicitly, we define the injected and post-decoding excess energies (relative to the ground state) by} 
	{$E_A:=\sum_k p_k\big(\langle \psi_k |H|\psi_k\rangle -E_0\big),\quad
	E_B:=\sum_k p_k\big(\langle \psi_k |(U_k^{B})^\dagger H U_k^{B}|\psi_k\rangle -E_0\big)$}
	
{In the nontrivial case where the measurement disturbs the ground state, one has $E_A>0$. If the measurement acts trivially on $|G\rangle$ (for example, all $M_k^{A}$ are proportional to $\mathbb{I}_A$ on $|G\rangle$), then $E_A=E_{\mathrm{ext}}=0$ and the protocol does nothing.}

To isolate the relevant terms, we utilize the elementary operator identity
\begin{equation}\label{eq:unitary_identity}
	(U_k^{B})^\dagger H U_k^{B} = H + (U_k^{B})^\dagger [H, U_k^{B}].
\end{equation}
Substituting this into the expression for $E_{\text{fin}}$, and using the property that $H|G\rangle = 0$, we obtain:
\begin{equation}
	\label{eq:energy_expansion}
	E_{\text{fin}} = \sum_k \langle G | (M_k^{A})^\dagger \left( H + (U_k^{B})^\dagger [H, U_k^{B}] \right) M_k^{A} | G \rangle.
\end{equation}
The first term, $E_A = \sum_k \langle G | (M_k^{A})^\dagger H M_k^{A} | G \rangle$, represents the average energy injected by Alice's measurement. This quantity is {non-negative}, and is strictly positive whenever the measurement is nontrivial on $|G\rangle$.

Crucially, due to the locality of the Hamiltonian, the commutator $[H, U_k^{B}]$ is supported only on sites within distance $R$ of region $B$. {More precisely, if $N_R(B):=\{x\in\Lambda:\operatorname{dist}(x,B)\le R\}$ denotes the $R$-neighbourhood, then $[H,U_k^{B}]$ is supported on $N_R(B)$ because only those interaction terms $h_Z$ with $Z\cap B\neq\varnothing$ fail to commute with $U_k^{B}$.} Let us define the local operator $O_k^{B} \equiv (U_k^{B})^\dagger [H, U_k^{B}]$. We can rewrite the total final energy as:
\begin{equation}
	E_{\text{fin}} = E_A + \sum_k \langle G | (M_k^{A})^\dagger O_k^{B} M_k^{A} | G \rangle.\label{eq:5}
\end{equation}

Since $M_a$ is supported on $A$ and $O_a^B$ is supported on the vicinity of $B$ i.e., $N_R(B)$, the cross-term $\langle G | M_a^\dagger O_a^B M_a | G \rangle$ is a ground-state correlation function. Specifically, expanding the density term $M_a |G\rangle\langle G| M_a^\dagger$, the ability of Bob to lower the energy depends entirely on the correlations between the local operators $\{M_a\}$ at $A$ and the energy flux operators $\{O_a^B\}$ at $B$.

For the protocol to successfully extract energy, this correlation term must take a negative value and our primary goal is to find nontrivial absolute bounds on this correlation term. To proceed, we analyze the energy difference using a quantum information-theoretic formalism.

\section{Spectral Constraints and Bounds}\label{sec:model} 
We continue with the model defined in Sec.~\ref{sec:framework} on the finite lattice $\Lambda$. {For any region $X\subset\Lambda$, we write $X^c:=\Lambda\setminus X$ for its complement.} We retain the assumptions regarding the Hamiltonian $H = \sum_{X \subset \Lambda} h_X$, specifically the uniform operator-norm bound $\|h_X\| \leq J$ and the finite interaction range {$R$}. To derive specific bounds on the energy extraction, we introduce an additional spectral constraint. We assume the Hamiltonian has a spectral gap $\Delta > 0$ separating the unique ground state energy $E_0=0$ from the first excited sector. We denote the ground-state density operator as $\omega = |G\rangle\langle G|$.

Alice performs a POVM with Kraus operators $\{M_a^{A}\}_a$ supported only on a bounded region $A\subset\Lambda$. The Kraus operators satisfy $\sum_a (M_a^{A})^\dagger M_a^{A}=\mathbb{I}_A$ and we assume the uniform operator-norm bound $\|M_a^{A}\|\le m$ for all $a$. We also assume outcome probabilities
$$
p_a:=\operatorname{tr}\big[((M_a^{A})^\dagger M_a^{A}\otimes \mathbb{I}_{A^c})\,\omega\big]>0,
$$
have a strictly positive minimum $p_{\min}>0$ (on a finite lattice this holds provided no Kraus operator corresponds to a zero Born probability). After the outcome $a$ is communicated to Bob, he applies a unitary $U_a^{B}$ supported only on region $B$, where $B$ is disjoint from $A$ and $d=\operatorname{dist}(A,B)>0$ denotes the separation. 

Energetic bookkeeping: the normalized post-measurement state for outcome $a$ is
\begin{equation}\label{eq:rho_a}
	\rho_a=\frac{(M_a^{A}\otimes\mathbb{I}_{A^c})\,\omega\,((M_a^{A})^\dagger\otimes\mathbb{I}_{A^c})}{p_a},
\end{equation}
and after Bob's unitary the state is
\begin{equation}\label{eq:sigma_a}
	\sigma_a=(U_a^{B}\otimes\mathbb{I}_{B^c})\,\rho_a\,((U_a^{B})^\dagger\otimes\mathbb{I}_{B^c}).
\end{equation}
Define the average excess energies (relative to $E_0$)
\begin{equation}\label{eq:E_AE_B}
	E_A:=\sum_a p_a\big(\operatorname{tr}[H\,\rho_a]-E_0\big),\qquad E_B:=\sum_a p_a\big(\operatorname{tr}[H\,\sigma_a]-E_0\big),
\end{equation}
We will nontrivially bound the net energy extraction $E_\text{ext}:=E_A-E_B$ from the above.

\subsection{Variational reduction}\label{sec:variational}
Fix an outcome $a$. By cyclicity of the trace and unitary invariance,
\begin{equation}\label{eq:deltaE_a}
	E_{\mathrm{ext}}^{(a)}:=\operatorname{tr}[H\,\rho_a]-\operatorname{tr}[H\,\sigma_a]=\operatorname{tr}\big[(H-(U_a^{B})^\dagger H U_a^{B})\,\rho_a\big].
\end{equation}
Insert the ground-state density $\omega$ and split linearly:
\begin{equation}\label{eq:split}
	E_{\mathrm{ext}}^{(a)}=\operatorname{tr}\big[(H-(U_a^{B})^\dagger H U_a^{B})(\rho_a-\omega)\big]+\operatorname{tr}\big[(H-(U_a^{B})^\dagger H U_a^{B})\,\omega\big].
\end{equation}
The second term is non-positive because the ground state minimizes energy: for any density operator $\tau$ we have $\operatorname{tr}[H\,\tau]\ge E_0$, and applying this to $\tau=U_a^{B}\omega (U_a^{B})^\dagger$ yields
$$
\operatorname{tr}\big[((U_a^{B})^\dagger H U_a^{B} - H)\,\omega\big]=\operatorname{tr}[H\,U_a^{B}\omega (U_a^{B})^\dagger]-\operatorname{tr}[H\,\omega]\ge0.
$$
Hence $\operatorname{tr}[(H-(U_a^{B})^\dagger H U_a^{B})\,\omega]\le0$, and dropping this non-positive term gives the variational upper bound
\begin{equation}\label{eq:varbound}
	\boxed{\;E_{\mathrm{ext}}^{(a)}\le \operatorname{tr}\big[(H-(U_a^{B})^\dagger H U_a^{B})(\rho_a-\omega)\big].\;}
\end{equation}
This reduces the problem to bounding the conjugation error $H-(U_a^{B})^\dagger H U_a^{B}$ and the local trace-norm $\|(\rho_a-\omega)\|_1$.

\section{Local decomposition and norm inequalities}\label{sec:local}
Decompose $H$ into local terms $h_X$ as in Sec.~\ref{sec:model}. Since $U_a^{B}$ acts nontrivially only on $B$, any term with $X\cap B=\varnothing$ satisfies $(U_a^{B})^\dagger h_X U_a^{B}=h_X$. Therefore
$$
(U_a^{B})^\dagger H U_a^{B} - H = \sum_{X:\;X\cap B\neq\varnothing} \big((U_a^{B})^\dagger h_X U_a^{B} - h_X\big).
$$
{Insert this into \eqref{eq:varbound} and expand the trace term-by-term:
	\begin{align}
		E_{\mathrm{ext}}^{(a)}
		&\le \sum_{X:\;X\cap B\neq\varnothing} \tr\!\Big[\big(h_X-U_a^\dagger h_X U_a\big)\,(\rho_a-\omega)\Big]\nonumber\\
		&=\sum_{X:\;X\cap B\neq\varnothing} \tr\!\Big[\big(h_X-U_a^\dagger h_X U_a\big)\,\big((\rho_a)_X-(\omega)_X\big)\Big]\nonumber\\
		&\le \sum_{X:\;X\cap B\neq\varnothing} \Big|\tr\!\Big[\big(h_X-U_a^\dagger h_X U_a\big)\,\big((\rho_a)_X-(\omega)_X\big)\Big]\Big| \nonumber\\
		&\le \sum_{X:\;X\cap B\neq\varnothing} \norm{h_X-U_a^\dagger h_X U_a}\;\norm{(\rho_a)_X-(\omega)_X}_1
		\label{eq:tracebound_pre}
	\end{align}
	where in the second line we used that $h_X-U_a^\dagger h_X U_a$ is supported on $X$, hence its expectation depends only on the reduced states on $X$. The second last line follows from triangle inequality and the last line from trace duality $|\tr[AB]|\le \norm{A}\,\norm{B}_1$. This implies the bound
	\begin{equation}\label{eq:tracebound}
		E_{\mathrm{ext}}^{(a)}\le \sum_{X:\;X\cap B\neq\varnothing} \norm{U_a^\dagger h_X U_a - h_X}\;\norm{(\rho_a-\omega)_X}_1
	\end{equation}
	where $(\cdot)_X$ denotes the reduction to $X$}

{To pass from \eqref{eq:tracebound} to the crude bound below we use, term-by-term,}
\begin{equation}\label{eq:triangle_step}
	{\|(U_a^{B})^\dagger h_X U_a^{B} - h_X\|\le \|(U_a^{B})^\dagger h_X U_a^{B}\|+\|h_X\|=2\|h_X\|\le 2J}
\end{equation}
Using \eqref{eq:triangle_step} yields
\begin{equation}\label{eq:crude}
	E_{\mathrm{ext}}^{(a)}\le 2\sum_{X:\;X\cap B\neq\varnothing} \|h_X\|\;\|(\rho_a-\omega)_X\|_1.
\end{equation}
Thus it remains to estimate the local trace-norms $\|(\rho_a-\omega)_X\|_1$ for $X$ that intersect $B$.

\subsection{Exponential clustering and local trace-norm estimates}\label{sec:clustering}
We employ the standard exponential clustering property for unique gapped ground states. A convenient form is the following lemma.

\begin{lemma}[Exponential clustering]\label{lem:cluster}
	Let $H$ be a finite-range Hamiltonian with a unique ground state $\omega$ and spectral gap $\Delta>0$. There exist constants {$c=c(\Delta,J,R)$} and {$\xi=\xi(\Delta,J,R)$} such that for any bounded operators $O_Y$ supported on $Y$ and $O_Z$ supported on $Z$,
	\begin{equation}\label{eq:cluster}
		\big|\langle G|O_Y O_Z|G\rangle - \langle G|O_Y|G\rangle\langle G|O_Z|G\rangle\big|\le c\,\|O_Y\|\,\|O_Z\|\,e^{-\operatorname{dist}(Y,Z)/\xi}
	\end{equation}
\end{lemma}

{A constructive choice of the correlation length follows from Nachtergaele--Sims:
for any $\lambda>0$ one may take $\xi=1/\mu$ with
$\mu=\frac{\Delta\lambda}{4\|\Phi\|_\lambda+\Delta}$, and $c=\mathcal{O}(1/\sqrt{\mu d})$,  where $\|\Phi\|_\lambda$ is the interaction norm defined in Eq.~(1) of \cite{NachtergaeleSims2006}. The correlation obtained here can be very loose compared to the specific model. This is understood as a consequence of using a uniform, model-independent bound: the Lieb--Robinson input is expressed through the global interaction norm $\|\Phi\|_\lambda$, which controls all local terms without exploiting any additional structure of the Hamiltonian. As a result, the derived $\mu$ (and hence $\xi$) reflects conservative estimates and need not be close to the actual correlation length of a given system;} see \cite{HastingsKoma2006,NachtergaeleSims2006,nachtergaele2006propagation,Lieb:1972wy} for constructive derivations.

To obtain a trace-norm bound for $\rho_{a,X}-\omega_X$ fix an outcome $a$, let $O_X$ be an arbitrary operator supported on $X$ with $\|O_X\|\le1$, and compute
\begin{align}
	\operatorname{tr}\big[O_X(\rho_{a,X}-\omega_X)\big]
	&=\frac{1}{p_a}\operatorname{tr}\big[((M_a^{A})^\dagger M_a^{A}\otimes O_X)\,\omega\big]-\operatorname{tr}\big[(\mathbb{I}_A\otimes O_X)\,\omega\big]\nonumber\\
	&=\frac{1}{p_a}\langle G|\big((M_a^{A})^\dagger M_a^{A} - p_a\mathbb{I}_A\big)\otimes O_X|G\rangle
\end{align}
Applying the clustering bound \eqref{eq:cluster} with $O_{Y=A}=(M_a^{A})^\dagger M_a^{A}$ and $O_Z=O_X$ and using $\|O_X\|\le1$ gives
$$
\big|\operatorname{tr}\big[O_X(\rho_{a,X}-\omega_X)\big]\big|\le \frac{c}{p_a}\,\|(M_a^{A})^\dagger M_a^{A}\|\,e^{-\operatorname{dist}(A,X)/\xi}
$$
Taking the supremum over $\|O_X\|\le1$ yields the explicit local trace-norm estimate
\begin{equation}\label{eq:trace_decay}
	\boxed{\;\|(\rho_a-\omega)_X\|_1 \le \frac{c\,\|(M_a^{A})^\dagger M_a^{A}\|}{p_a}\;e^{-\operatorname{dist}(A,X)/\xi}.\;}
\end{equation}

\subsection{Main bound}\label{sec:main}
Insert \eqref{eq:trace_decay} into \eqref{eq:crude} to obtain for each $a$:
$$
E_{\mathrm{ext}}^{(a)}\le 2\sum_{X:\;X\cap B\neq\varnothing} \|h_X\|\;\frac{c\,\|(M_a^{A})^\dagger M_a^{A}\|}{p_a}\;e^{-\operatorname{dist}(A,X)/\xi}.
$$
Averaging over $a$ with weights $p_a$ and interchanging sums gives
\begin{equation}\label{eq:DeltaE_avg}
	E_{\mathrm{ext}}=\sum_a p_a E_{\mathrm{ext}}^{(a)}\le 2c\sum_a\|(M_a^{A})^\dagger M_a^{A}\|\;\sum_{X:\;X\cap B\neq\varnothing} \|h_X\|\;e^{-\operatorname{dist}(A,X)/\xi}.
\end{equation}
Define the boundary sums
\begin{equation}\label{eq:boundary_sums}
	{S_B:=\sum_{X:\;X\cap B\neq\varnothing} \|h_X\|,\qquad S_A:=\sum_a\|(M_a^{A})^\dagger M_a^{A}\|.}
\end{equation}
{The restriction $X\cap B\neq\varnothing$ arises because only Hamiltonian terms intersecting Bob's support can change under conjugation by $U_a^B$; equivalently, the sum ranges over the $R$-thick boundary layer around $B$.}

For any $X$ with $X\cap B\neq\varnothing$ we have $\operatorname{dist}(A,X)\ge d-\operatorname{diam}(X)\ge d-R$, hence $e^{-\operatorname{dist}(A,X)/\xi}\le e^{R/\xi}e^{-d/\xi}$. Using this bound in \eqref{eq:DeltaE_avg} yields
\begin{equation}\label{eq:final}
	\boxed{\;E_{\mathrm{ext}}\le C\,e^{-\mu d},\qquad C:=2c\,S_A\,S_B\,e^{R/\xi},\quad \mu:=1/\xi.\;}
\end{equation}

\section{Refinement: commutator control of decoding unitaries}\label{sec:comm}
The crude operator-norm bound used above can be sharpened when each decoding unitary admits a local self-adjoint generator. {Since $U_a^{B}$ acts on a finite-dimensional Hilbert space (the degrees of freedom confined in the region $B$), one can diagonalize $U_a^{B}=V\operatorname{diag}(e^{i\theta_j})V^\dagger$ with $\theta_j\in(-\pi,\pi]$ and define $G_a:=V\operatorname{diag}(\theta_j)V^\dagger$, so that $G_a=G_a^\dagger$ and $U_a^{B}=e^{iG_a}$.} Suppose $U_a^{B}=e^{iG_a}$ with $G_a=G_a^\dagger$. We define
$$
f(s):=e^{\,i s G_a}\,h_X\,e^{-\,i s G_a},\quad s\in\mathbb{R}.
$$
A direct computation yields
\begin{equation}
	\frac{d}{ds}f(s)
	= i\,e^{\,i s G_a}\,[G_a,h_X]\,e^{-\,i s G_a}.
	\label{eq:fs-derivative}
\end{equation}
Integrating \eqref{eq:fs-derivative} from $s=0$ to $s=1$ gives
$$
(U_a^{B})^\dagger h_X U_a^{B} - h_X = i\int_0^1 e^{-isG_a}[G_a,h_X]e^{isG_a}\,ds,
$$
and taking norms gives
$$
\|(U_a^{B})^\dagger h_X U_a^{B} - h_X\|\le \|[G_a,h_X]\|.
$$
Using this in place of the crude norm in \eqref{eq:tracebound} gives the refined bound
\begin{equation}\label{eq:refined}
	\boxed{\;E_{\mathrm{ext}}\le \tilde C\,e^{-\mu d},\quad \tilde C:=c\,e^{R/\xi}\sum_a\|(M_a^{A})^\dagger M_a^{A}\|\sum_{X:\;X\cap B\neq\varnothing}\|[G_a,h_X]\|.\;}
\end{equation}

\section{{Worked example and comparison with toy models}}\label{sec:example}
The theorem above is model-independent, so it is natural to ask how it connects to explicit QET constructions in the existing literature. We give a simple worked example in a gapped spin chain.

Consider a one dimensional nearest neighbor spin-$\tfrac12$ chain with Hamiltonian $H=\sum_{i}h_{i,i+1}$ where $\norm{h_{i,i+1}}\le J$ and the unique ground state is separated from the rest of the spectrum by a gap $\Delta>0$. Let $A=\{i\}$ and $B=\{j\}$ with separation $d=|i-j|$. Take Alice's projective measurement in the $\sigma_i^z$ basis,
$$
M_0=\frac{\mathbb{I}+\sigma_i^z}{2},\qquad M_1=\frac{\mathbb{I}-\sigma_i^z}{2},
$$
and let Bob apply an outcome-dependent local rotation
$$
U_a=\exp\!\big(i(-1)^a\theta\,\sigma_j^y\big),\qquad a\in\{0,1\}.
$$
For nearest-neighbor interactions and single-site $B$, the neighborhood Hamiltonian is $h_{j-1,j}+h_{j,j+1}$, hence $S_B\le 2J$. Moreover $\norm{M_a^\dagger M_a}=1$ so $S_A=2$. Therefore our general bound \eqref{eq:final} yields
\begin{equation}\label{worked_example_ext}
	 E_\text{ext} \le 8c\,J\,e^{R/\xi}\,e^{-d/\xi}
\end{equation}
This captures the qualitative behavior of QET protocols for gapped chains in the gapped regime.

\paragraph{Toy model: transverse-field Ising chain and physical correlation length.}
As a benchmark, consider Pfeuty's transverse-field Ising chain \cite{Pfeuty1970}
\begin{equation}\label{TFIM}
H=-r\sum_\ell S_\ell^z - J\sum_\ell S_\ell^x S_{\ell+1}^x
\end{equation}
which has an unique, entangled and gapped ground state for $r> J/2$ with the gap
$
\Delta=\left|r-\frac{J}{2}\right|.
$
In this gapped paramagnetic phase $r>J/2$ (equivalently $\lambda=J/(2r)<1$), the equal-time correlator obeys the large-distance asymptotic
$p_n^{x}\sim \frac14(1-\lambda^{2})^{-1/4}\pi^{-1/2}n^{-1/2}\lambda^{n}$ (Pfeuty, Eq.~(3.8)).
Hence $\lambda^{n}=e^{-n/\xi_x}$ and the correlation length is $\xi_x=[-\ln\lambda]^{-1}=[\ln(2r/J)]^{-1}$ \cite{Pfeuty1970}.

Whereas, one can use the Lemma \ref{lem:cluster} to obtain clustering correlation length associated to the Hamiltonian \ref{TFIM}. Writing $H=\sum_X h_X$ with
$$
h_{\{\ell\}}=-rS_\ell^z,\qquad h_{\{\ell, \ell + 1\}}=-J S_\ell^x S_{\ell+1}^x
$$
and using $\|S^\alpha\|=\tfrac12$, one has
$
\|h_{\{\ell\}}\|=\frac{|r|}{2},\quad \|h_{\{\ell, \ell + 1\}}\|=\frac{|J|}{4}
$.
For the weighted interaction norm $\|\Phi\|_\lambda$ used in \cite{NachtergaeleSims2006}, consequently with our notation, $\|h\|_\lambda$, a direct counting of the on-site term and the two adjacent bonds gives
$
\|h\|_\lambda \le 2|r|+16|J|e^\lambda,\qquad \lambda>0
$
The corresponding clustering rate can then be taken as \cite{NachtergaeleSims2006}
$$
\mu(\lambda)=\frac{\lambda \, \Delta}{4\|h\|_\lambda+\Delta}
\ge \frac{\lambda \, \Delta}{(8|r|+\Delta)+64|J|e^\lambda},
\qquad
\xi(\lambda)=\mu(\lambda)^{-1}
$$
Optimizing over $\lambda$ yields an explicit choice. Setting $P:=8|r|+\Delta$ and $Q:=64|J|$, the maximizer satisfies
$
\lambda_\ast = 1 + W\!\left(\frac{P}{Qe}\right)
$,
where $W$ is the Lambert $W$-function, and
\begin{equation}
	\xi_\ast=\xi(\lambda_\ast)
	=\frac{P}{\Delta\,W\!\left(\frac{P}{Qe}\right)}
	=\frac{8|r|+\Delta}{\Delta\,W\!\left(\frac{8|r|+\Delta}{64e|J|}\right)}
\end{equation}
The length $\xi_\ast$ applies broadly, but it is typically conservative compared to the physical correlation length $\xi_x$ extracted from model such as \cite{Pfeuty1970}. This explains why \eqref{eq:final} correctly captures exponential suppression with distance in gapped regimes, while generally overestimating the true decay length.

\section{Conclusions}
The bounds \eqref{eq:final} and \eqref{eq:refined} are non-perturbative and explicit. Given numerical values of the gap $\Delta$, the local-norm bound $J$, the interaction range $R$, and the POVM parameters, one can (in principle) determine the decay constants \(c,\xi\) --- for instance, (non-optimally) from the proofs of exponential clustering \cite{NachtergaeleSims2006} --- and hence evaluate \(C,\tilde{C}\) explicitly. However, carrying out these computations for concrete models is technical. {Nevertheless, we presented a detailed derivation of a bound for a transverse-field Ising chain in the gapped regime. Moreover, we gave a benchmark for the correlation length in Section~\ref{sec:example}.}

The essential takeaway is the following.  {In gapped systems (in thermodynamic sense) with a unique ground state, our result gives an explicit energetic consequence of exponential clustering: the local disturbance on correlation created by Alice decays exponentially in trace norm with distance, which forces the QET-extractable energy to be exponentially small at large separation. Consequently, a spatially separated agent who receives only classical information can extract at most an amount of energy that is exponentially small in the separation by performing local operations on $B$.}

Possible extensions include (i) explicit evaluation of the constants $c,\xi$ for various specific known models using constructive Lieb--Robinson estimates, and (ii) extending the bound to thermal states with sufficiently low temperature (using clustering for thermal states).

\bibliographystyle{elsarticle-num}

\end{document}